\documentclass[a4paper]{jpconf}

\usepackage[square,sort&compress, numbers]{natbib}
\usepackage{graphicx}
\usepackage[utf8]{inputenc}
\usepackage[T1]{fontenc}

\graphicspath{{figures/}}
\DeclareGraphicsExtensions{.pdf, .eps}

\newcommand{\tsup}[1]{\textsuperscript{#1}}
\newcommand{\molec}[2]{#1$_{#2}$}
\newcommand{\zco}{ \molec{Zn}{}\molec{Cr}{2}\molec{O}{4}}
\newcommand{\mco}{ \molec{Mg}{}\molec{Cr}{2}\molec{O}{4}}
\newcommand{\cco}{ \molec{Cd}{}\molec{Cr}{2}\molec{O}{4}}

\begin{document}

\title{Magnetic susceptibility of the frustrated spinels \zco{}, \mco{} and \cco{}}

\author{Ch. Kant,\tsup{1}  J. Deisenhofer,\tsup{1} V. Tsurkan\tsup{1,2} and A. Loidl\tsup{1}}

\address{\tsup{1}\ Experimental Physics~V, Center for Electronic
Correlations and Magnetism, University of Augsburg, D-86135~Augsburg,
Germany}
\address{\tsup{2}\ Institute of Applied Physics, Academy of Sciences of Moldova,
MD-2028~Chişinău, Republic of Moldova}

\ead{christian.kant@physik.uni-augsburg.de}

\begin{abstract}
We analyzed the magnetic susceptibilities of several Cr spinels using two recent models for the geometrically frustrated pyrochlore lattice, the Quantum Tetrahedral Mean Field model and a Generalized Constant Coupling model. Both models can describe the experimental data for $A$\molec{Cr}{2}\molec{O}{4} (with $A$ = Zn, Mg, and Cd) satisfactorily, with the former yielding a somewhat better agreement with experiment for $A$ = Zn, Mg.  The obtained exchange constants for nearest and next-nearest neighbors are discussed.
\end{abstract}

%\section{Introduction}
The spinel systems $A$\molec{Cr}{2}\molec{O}{4}, where $A$ is non-magnetic, are a paradigm for highly frustrated lattices. The Cr\tsup{3+} ions with spin $s=3/2$ form a network of corner sharing tetrahedra which is isomorphic to the pyrochlore lattice. For classical Heisenberg spins with antiferromagnetic (AF) nearest-neighbor (nn) interaction, it was predicted that these systems do not order until lowest temperatures \cite{moessner98a}. If next-nearest neighbor (nnn) exchange is taken into account, however, the huge ground-state degeneracy is lifted and magnetic order sets in when approaching zero temperature \citep{chern08}.  In reality, the Cr oxide-spinels were reported to exhibit AF ordering with Néel temperatures of 12.5, 12.7 and 7.8 \cite{rovers02,lee00} for \zco{}, \mco{} and \cco{}, respectively, albeit the respective Curie-Weiss temperatures are -390, -346 and -71~K \cite{rudolf07a,sushkov05}. This transition is accompanied by a structural distortion evidencing the importance of magnetoelastic coupling as a means to relieve magnetic frustration \cite{tchernys02,sushkov05,rudolf07a,aguilar08}. Garcia-Adeva and Huber put forward two models to describe the temperature dependence of the magnetic susceptibility of frustrated pyrochlore paramagnets which allow to estimate the nn and nnn exchange coupling \cite{garcia-a00,garcia-a02b}. These models were tested for \zco{} and yielded reasonable results \cite{martinho01, garcia-a02b}, for \mco{} and \cco{} we are not aware of any detailed investigations.

The first model, the Quantum Tetrahedral Mean Field (TMF) model \cite{garcia-a00} is used in the form
\begin{equation}\label{eq:chiTMF}
\chi_{TMF}(T) = \frac{N_A g^2 \mu_B^2}{k_B} \frac{a \cdot \chi_{tet}(T)}{1+(3 J_1 +12 J_2)\chi_{tet}(T)}, 
\end{equation}
which describes the susceptibility per mole of magnetic ions. Here, $N_A$ is the Avogadro constant,  $g=1.97$ the $g$-factor \cite{martinho01}, $\mu_B$ the Bohr magneton, and $J_1$
and $J_2$ are the nn and nnn exchange constants (in units of $k_B$), respectively. $\chi_{tet}$ is given by
\begin{equation}\label{eq;chiTet}
\chi_{tet}(T) = \frac{1}{12 T} \frac{\sum\limits_S g(S) S (S+1)
(2S+1) e^{\frac{-J_1 S (S+1)}{2 T}}} {\sum\limits_S g(S) (2S+1)
e^{\frac{-J_1 S (S+1)}{2 T}}}.
\end{equation}
The sum runs over the total spin values $S=(0,1,2,3,4,5,6)$ of the Cr-tetrahedron and $g(S)=(4,9,11,10,6,3,1)$ are the corresponding degeneracy factors \cite{garcia-a00}. The scaling factor $a$ is related to the effective number of Bohr magnetons by $a=p^2_{eff} / [g^2 s(s+1)]$.

The second model is a generalization of the constant coupling model (Refs.~\cite{garcia-a02b,garcia-a02} and references therein). We are using the most recent results \cite{garcia-a02b}. Here, the magnetic susceptibility per mole of magnetic ions is given by
\begin{equation}\label{eq:chiGCC}
\chi_{GCC}(T) = \frac{N_A g^2 \mu_B^2}{k_B} \cdot \chi_1(T) \cdot \frac{a\cdot(1+\varepsilon(T))}{1-\varepsilon(T)\left( 1+ 4\; J_2/J_1\right)},
\end{equation}
where $\varepsilon(T)$ is defined by $\varepsilon(T)=\chi_{tet}/\chi_1$ and $\chi_1(T)$ by $\chi_1(T)=s(s+1)/3T$.

Both models share common roots therein that they first focus on an isolated tetrahedra with only nn interaction and then incorporate nnn couplings. In the first model this is done in a heuristic manner by calculating the interaction with ions outside the tetrahedron with an effective coupling constant, and in a self-consistent relation of the internal magnetic field in the latter one.

%\section{Results and discussion}

\begin{figure}
\center
\includegraphics[width=0.9\textwidth]{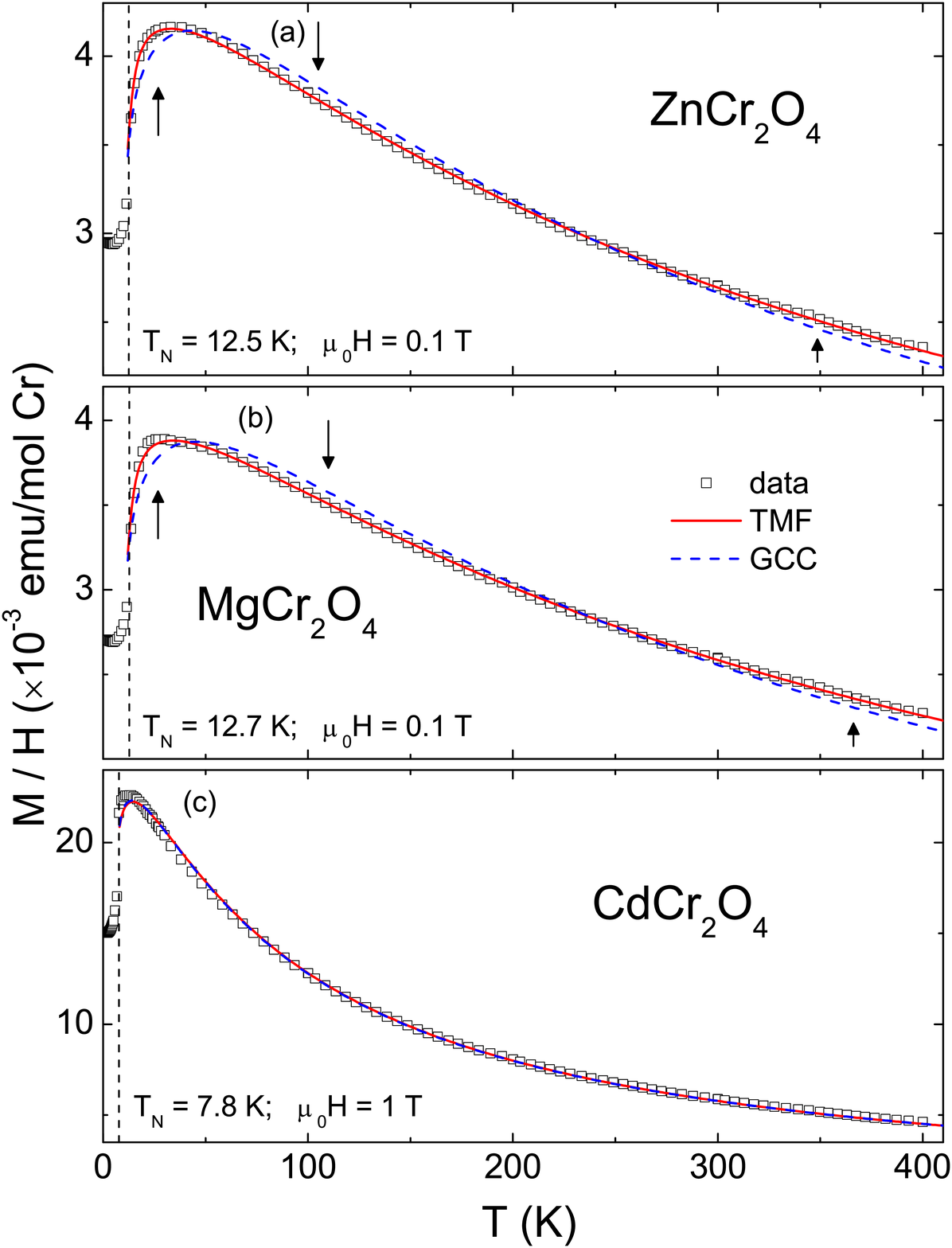}
\caption{\label{fig:chiACO} Temperature dependencies of the magnetic susceptibilities of (a) \zco{}, (b) \mco{} and (c) \cco{}. Also shown are the best fits according to the TMF (solid lines) and GCC (dashed lines) model. Arrows indicate clear deviations from the data.}
\end{figure}

Figure~\ref{fig:chiACO} shows the magnetic susceptibility of single crystalline samples of all three Cr oxides which were grown by chemical transport or flux method. The quality was checked by x-ray diffraction. The susceptibility measurements were carried out in a commercial SQUID (Quantum Design). No evident differences in the susceptibility $\chi$ were detected between poly- and single crystals in the paramagnetic state.

The fit curves according to Eq.~(\ref{eq:chiTMF}) and (\ref{eq:chiGCC}) are also shown in Fig.~\ref{fig:chiACO}. The corresponding fit parameters are listed in Table~\ref{tab:params}. In the case of \zco{} and \mco{} the TMF model fits the data in the whole temperature range above $T_N$ very nicely, yielding effective moments of 3.69~$\mu_B$ and 3.70~$\mu_B$, respectively. These values are in good agreement with the spin-only value of $p_{eff}=3.87 \mu_B$ for Cr$^{3+}$. In contrast, the best fit with the GCC model exhibits clear deviations from the data (indicated by arrows) and results in lower effective moments of 3.44~$\mu_B$ and 3.41~$\mu_B$. In addition, the obtained nn and nnn exchange constants $J_1$ and $J_2$ differ considerably for these two models. For \zco{} a similar discrepancy of these two models was reported previously Ref.~\cite{martinho01}. The TMF model yielded $J_1=39.4$~K and $J_2=1.76$~K while the application of the GCC model resulted in $J_1=27.23$~K and $J_2=3.00$~K \cite{garcia-a02b}.

For \cco{}, where the antiferromagnetic nn exchange is already weakened in comparison to the other two compounds and the nnn exchange is ferromagnetic, both models yield almost identical curves and fitting parameters $p_{eff}\simeq 4.0 \mu_B$, $J_1\simeq15$~K, and $J_1\simeq -4$~K. This decrease of the antiferromagnetic nn coupling and the occurrence of ferromangetic nnn exchange is in agreement with the phase diagram of the Cr spinels \cite{rudolf07a}.

Despite a considerable number of studies on these frustrated magnets, independently determined values for $J_1$ and $J_2$ are not easy to find in the literature. Optical and ESR studies of Cr-Cr pairs in ZnGa$_2$O$_4$ yielded $J_1=32$~K \cite{henning73,vangorko73}, which is a little closer to the results of the TMF model. ESR studies of \zco{} derived in a similar way resulted in $J_1=45$~K \cite{martinho01}, again closer to the TMF result and in agreement with the estimate using the high-temperature CW temperature via $J_1=-3k_B\theta_{CW}/z S(S+1)$=52~K, where $z=6$ is the number of nearest neighbors and $\theta_{CW}$=-390~K  \cite{ueda06}. The corresponding values of $J_1$ are 46~K and 9.5~K (using CW temperatures of -346~K and -71~K) for \mco{} and \cco{}, respectively.

Therefore, we conclude that for the strongly antiferromagnetically coupled systems \zco{} and \mco{} the TMF model yields a better fit of the data, while for \cco{} the GCC approach is similarly good. Additionally, we would like to point out that none of the two models allowed for a reasonable fitting of the magnetic susceptibilites of the Cr spinels ZnCr$_2$S$_4$ and ZnCr$_2$Se$_4$, where the nnn FM exchange becomes even stronger than in \cco{} \cite{rudolf07a}.

\begin{table}
\center
\caption{\label{tab:params}Parameters obtained by fitting the TMF and GCC models to the experimental data.}
\begin{tabular}{c@{\hspace{4ex}}*{8}{c}}
\br
&\multicolumn{2}{c}{\zco{}}&&\multicolumn{2}{c}{\mco{}}&&\multicolumn{2}{c}{\cco{}}\\[0.1ex]
\cline{2-3}\cline{5-6}\cline{8-9}\\[-2.2ex]
&TMF&GCC&&TMF&GCC&&TMF&GCC\\
\mr
$J_1/\textrm{K}$  &  33.4  &  25.1  &&  34.4  & 25.5  &&  14.7  &  15.2\\
$J_2$/\textrm{K}  &  4.4  &  2.7  &&  5.7  &  3.4  &&  -4.0  &  -4.1\\
$p_{eff}/\mu_B$  &  3.69  &  3.44  &&  3.70  &  3.41  &&  4.04  &  4.05\\
\br
\end{tabular}
\end{table}

\ack We thank D. L. Huber and A. J. García-Adeva for fruitful
discussions and D. Vieweg for technical assistance. This work was
partly supported by the Deutsche Forschungsgemeinschaft DFG by the
Collaborative Research Center SFB~484 (University of Augsburg).

\providecommand{\newblock}{}

\end{document}